\documentclass[smallextended]{svjour3}       
\smartqed  
\usepackage{graphicx}
\begin{document}

\title{Kaons production at finite temperature and baryon density in an effective relativistic mean field model}

\author{F. Iazzi         \and
        A. Lavagno        \and
        D. Pigato
}
\institute{Dipartimento di Fisica, Politecnico di Torino, I- 10129 Torino, Italy\\
           Istituto Nazionale di Fisica Nucleare (INFN), Sezione di Torino, I-10126 Torino, Italy
}

\date{Received: date / Accepted: date}

\vspace{-1cm}
\maketitle

\begin{abstract}
We investigate the kaons production at finite temperature and baryon
density by means of an effective relativistic mean-field model with the inclusion
of the full octet of baryons. Kaons are considered taking
into account of an effective chemical potential depending on the self-consistent
interaction between baryons. The obtained results are compared with a minimal coupling
scheme, calculated for different values of the anti-kaon optical potential.
\end{abstract}

\section{Hadronic equation of state and main results}
The relativistic mean-field model (RMF) is widely successful used for describing the
properties of finite nuclei as well as hot and dense nuclear matter
\cite{1,2,3,glen0,4,prl,epj,plb}.
In this context, the total baryon Lagrangian density can be written as
${\cal L}_B={\cal L}_{\rm octet}+{\cal L}_K$,
where ${\cal L}_{\rm octet}$ stands for the full octet of baryons
($p$, $n$, $\Lambda$, $\Sigma^+$, $\Sigma^0$, $\Sigma^-$, $\Xi^0$,
$\Xi^-$) and ${\cal L}_K$ corresponds to the kaon mesons.

The quantum hadrodynamics (QHD) model for the full octet of baryons was
originally studied with the following standard Lagrangian density \cite{glen0}
\begin{eqnarray}\label{lagrangian}
{\cal L}_{\rm octet} \!\!\!&=&\!\!\!
\sum_i\bar{\psi}_i\,[i\,\gamma_{\mu}\,\partial^{\mu}-(m_i- g_{\sigma
i}\,\sigma) -g_{\omega
i}\,\gamma_\mu\,\omega^{\mu} -g_{\rho i}\,\gamma_{\mu}\,\vec{\tau}
\cdot \vec{\rho}^{\;\mu}]\,\psi_i \nonumber\\
&&+\,\frac{1}{2}(\partial_{\mu}\sigma\partial^{\mu}\sigma-m_{\sigma}^2\sigma^2)
-U(\sigma)+\frac{1}{2}\,m^2_{\omega}\,\omega_{\mu}\omega^{\mu}
+\frac{1}{4}\,c\,(g_{\omega N}^2\,\omega_\mu\omega^\mu)^2\nonumber\\
&&+\,\frac{1}{2}\,m^2_{\rho}\,\vec{\rho}_{\mu}\cdot\vec{\rho}^{\;\mu}
-\frac{1}{4}F_{\mu\nu}F^{\mu\nu}
-\frac{1}{4}\vec{G}_{\mu\nu}\vec{G}^{\mu\nu}\, ,
\end{eqnarray}
where $m_i$ is the vacuum baryon mass of index $i$, $\vec{\tau}=2\,\vec{t}$ denotes the isospin
operator and $U(\sigma)$ is the nonlinear self-interaction potential of $\sigma$ meson \cite{2,4}.
It is proper to remark that the above QHD Lagrangian is very different from the genuine SU(3) models, defined in terms of baryon and meson octets \cite{gal_ref}.

Because we are going to describe finite temperature and density
nuclear matter with respect to strong interaction, we have to
require the conservation of three "charges": baryon
number, electric charge and strangeness number. Therefore, the chemical potential of particle
of index $i$ can be written as
\begin{equation}
\mu_i=b_i\, \mu_B+c_i\,\mu_C+s_i\,\mu_S  \, ,\label{mu}
\end{equation}
where $b_i$, $c_i$ and $s_i$ are, respectively, the baryon, the
electric charge and the strangeness quantum numbers of the $i$-th
hadrons. The effective chemical potential $\mu_i^*$ of the $i$-th baryon is given by
$\mu_i^*=\mu_i-g_{\omega i}\,\omega-g_{\rho i}\,\tau_{3 i}\,\rho$.
%

In this context, kaons degrees of freedom are treated in two distinct approaches. In the first case, we consider the interaction between kaons and baryons by means of a direct minimal coupling scheme with the meson fields \cite{5,6,7}. The kaon lagrangian density can be written as ${\mathcal L}_{K}= D_{\mu}^*\Phi^* D^{\mu}\Phi -m_{K}^{*2}\Phi^*\Phi$,
where $D_{\mu}=\partial_{\mu} +ig_{\omega K}\omega_{\mu} +ig_{\rho
K}\tau_{3 K}\rho_{\mu}$ is the covariant derivative of the meson field,
$m_{K}^*=m_K-g_{\sigma K}\sigma$ is the effective kaon mass and
$\tau_{3 K}$ is the third component of the isospin operator.
The kaon-meson vector coupling constants are obtained
from the quark model and isospin counting rules, setting equal to
$g_{\omega K}= g_{\omega N}/3$ and $g_{\rho K}=g_{\rho N}$.
Whereas the scalar $g_{\sigma K}$ coupling constant is determined from the
study of the real part of the anti-kaon optical potential, at
saturation nuclear density, in symmetric nuclear matter:
$U(K^-)= -g_{\sigma K}\sigma - g_{\omega K}\omega$. In this investigation we set the
anti-kaon optical potential equal to $U(K^-)=-50$ MeV, $-100$ MeV and $-160$ MeV,
based on recent theoretical calculations and experimental measurements \cite{9,10,11,12,13}.
The meson-baryons couplings constant have been fixed to the parameters set marked as GM3 of Ref.s \cite{4,8}.

In the second approach, we use an alternative formulation, based on the
self-consistent interaction between baryons \cite{8}. In this scheme, kaons are treated
as a quasi-ideal Bose gas with an effective chemical potential $\mu^*_K$, obtained from the "bare" one given in Eq.(\ref{mu}) and subsequently expressed in terms of the corresponding effective baryon chemical potentials, respecting the strong interaction. More explicitly, the kaon effective chemical potential can be written as
\begin{eqnarray}
&\mu^*_{K^+} = (\mu^*_p -\mu^*_{\Lambda}) = \mu_p -\mu_{\Lambda} - (1- x_{\omega \Lambda})g_{\omega N}\omega -\frac{1}{2}g_{\rho N}\rho\ ,
\end{eqnarray}
where $x_{\omega \Lambda}=g_{\omega \Lambda}/g_{\omega N}$.
Thus, the hadronic system is still regarded as
an ideal gas but with an effective chemical potential that contains the
self-consistent interaction of the meson fields, related to the interaction
between baryons.

In this paper we focus our attention into a comparative study between the above two different approaches. At this scope, for simplicity, other strangeless mesons (mainly pions) are not considered in our analysis, assuming that they do not sensibly affect the strangeness production but contribute essentially to the total pressure and energy density. Heavier strange meson degrees of freedom have been also neglected.

At finite temperature the thermodynamical quantities can be obtained in the standard way from the total grand potential $\Omega_H=\Omega_B +\Omega_K $ for the two different approaches \cite{6,7}.

In the Fig. 1, we report the major results obtained in this comparative study.
In both panels we observe a good correspondence between the effective relativistic mean-field model (solid lines) and the minimal coupling scheme (dashed lines) for moderate values of the anti-kaon optical potential. In particular, recent self-consistent calculations based on coupled-channel G-matrix theory \cite{9} and chiral Lagrangian \cite{14} seem to suggest that the real part of the anti-kaon optical potential could be close to $U_{K^-} \cong -50$ MeV, in good
agreement with our results.
\begin{figure}
\includegraphics[width=1.0\textwidth]{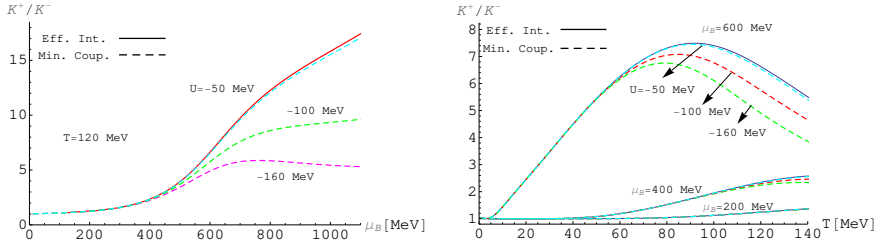}
\caption{Kaon to anti-kaon ratio as a function of baryon chemical potential (left panel) and temperature (right panel) in the effective relativistic mean field model (solid lines) and for different values of anti-kaon optical potential (dashed lines)}
\end{figure}
Moreover, it is interesting to observe that, for a very strong attractive optical
potential ($U_{K^-}=-160$ MeV) and for high values of $\mu_B$, a lowering in the $K^+/K^-$ ratio takes place, especially at high temperature. This behavior could be considered as a relevant feature for the determination of the real part of the anti-kaon optical potential, for example in the future CBM (compressed baryonic matter) experiment of the FAIR (Facility of Antiproton
and Ion Research) project \cite{senger}.


\begin{thebibliography}{}
\bibitem{1}
Walecka J.D.: Ann. of Phys. {\bf 83}, 491 (1974)
\bibitem{2}
Boguta J., Bodmer A.R.: Nucl. Phys. {\bf A292}, 413 (1977)
\bibitem{3}
Serot B.D., Walecka J.D.: Adv. Nucl. Phys. {\bf 16}, 1 (1986)
\bibitem{glen0}
Glendenning N.K.: Phys. Lett. {\bf B114}, 398 (1982)
\bibitem{4}
Glendenning N.K., Moszkowski S.A.: Phys. Rev. Lett. {\bf 67}, 2414 (1991)
\bibitem{prl}
Bonanno L., Drago A., Lavagno A.: Phys. Rev. Lett. {\bf 99}, 242301 (2007)
\bibitem{epj}
Alberico W.M., Lavagno A.: Eur. Phys. J. {\bf A40}, 313 (2009)
\bibitem{plb}
Lavagno A., Quarati P.: Phys. Lett. {\bf B498}, 47 (2001)
\bibitem{gal_ref}
Schaffner-Bielich J., Gal A.: Phys. Rev. {\bf C62}, 034311 (2002) and references therein
\bibitem{5}
Glendenning N.K., Schaffner-Bielich J.: Phys. Rev. {\bf C60}, 025803 (1999)
\bibitem{6}
Cavagnoli R., Providencia C., Menezes D.P.: Phys. Rev {\bf C83}, 045201 (2011)
\bibitem{7}
Banik S., Greiner W., Bandyopadhyay D.: Phys. Rev. {\bf C78}, 065804 (2008)
\bibitem{9}
Tol'os L., Ramos A., Polls A.: Phys. Rev. {\bf C65}, 054907 (2002)
\bibitem{10}
Friedman E., Gal A., Batty C.J.: Nucl. Phys. {\bf A579}, 518 (1994)
\bibitem{11}
Batty C.J., Friedman E., Gal A.: Phys. Rep. {\bf C60} 024314 (1999)
\bibitem{12}
Friedman E., Gal A., Mares J., Ciepl´y A.: Phys. Rev. {\bf C60}, 024314 (1999)
\bibitem{13}
Gal A.: Nucl. Phys. {\bf A691}, 268 (2001)
\bibitem{8}
Lavagno A.: Phys. Rev. {\bf C81}, 044909 (2010)
\bibitem{14}
Lutz M.F.M., Korpa C.L.: Nucl. Phys. {\bf A700}, 309 (2002)
\bibitem{senger}
Senger P.: J. Phys. G: Nucl. Part. Phys. {\bf 30}, S1087 (2004);
Senger P. et al.: ibid. {\bf 36}, 064037 (2009); Henning W.F.: Nucl.
Phys. {\bf A805}, 502c (2008)
\end{thebibliography}
\end{document}